\documentclass[a4paper]{jpconf}
\usepackage{graphicx}
\begin{document}
\title{The white dwarf in dwarf nova SDSS J080434.20+510349.2: Entering the instability strip?}

\author{E Pavlenko$^{1,2}$}

\address{$^1$Crimean astrophysical observatory, Crimea 98409, Ukraine}
\address{$^2$Tavrida National Vernadsky University, Crimea 95007, Ukraine}

\ead{pavlenko@crao.crimea.ua}

\begin{abstract}
SDSS J080434.20+510349.2 is the WZ type binary that displayed rare outburst in 2006 (Pavlenko et al., 2007). During the long-lasting tail of the late stage of the outburst binary shown the two-humped or four-humped profile of the orbital light modulation. The amplitude of orbital light curve decreased while the mean brightness decreased, more over that occurred $\sim$ 10 times faster during the fast outburst decline in respect to the late quiet state of slow outburst fading.  There were no white dwarf pulsations detected neither 1 - 1.5 months prior to the outburst nor in 1.5 - 2 months after the 2006 outburst in this system. However the strong non-radial pulsations with period 12.6 minutes and mean amplitude of $0.05^{m}$ were first detected in V band with 2.6-m Shajn mirror telescope of the Crimean astrophysical observatory in $\sim$ 8 months after the outburst. The evolution of pulsations over two years in 2006 - 2008 is considered. It is supposed that pulsations first appeared when the cooling white dwarf (after the outburst) entered the instability strip although the possibility of temporary lack of pulsations at some occasions also could not be excluded.
\end{abstract}

\section{Introduction}
12 cataclysmic variables (CVs) with orbital period close to the evolutional orbital period minimum, containing the pulsating white dwarf (WD), were known up to now (Mukadam et al., this volume). SDSS J080434.20+510349.2 (hereafter SDSS J0804) is the 13th  system showing such pulsations (Pavlenko and Malanushenko, 2009) which are generally believed are due to non radial g-mode pulsations of the white dwarf (Warner and Robinson, 1972).  P. Szkody et al. reported that there are 4 dwarf novae: PQ And, GW Lib, V455 And, REJ 1255+26 containing the accreting WD pulsator that undergone the outburst (is available from www.narit.or.th/conference/prcsa2008/).  SDSS J0804 (Pavlenko et al., 2007) is the 5th such binary.  SDSS J0804 is the WZ Sge type star displayed the outburst in 2006 with 11 rebrightenings. Superhumps that have been observed both during the main outburst and rebrightenings  later on were replaced by the orbital two-humped light variations (Zharikov et al., 2008; Pavlenko, 2007) typical to the WZ-Sge stars in quiescence. The phenomenon of a second series of rebrightenings or "`mini-outburst"' with longer recurrence time and less amplitude was found by Zharikov et al (2008). In quiescence before outburst optical spectrum of SDSS J0804 displayed a blue continuum with broad absorption lines from a white dwarf surrounding the double-peaked Balmer emission lines formed in accretion disk (Szkody et al., 2006). Authors also noted the particular behavior of the binary during one night in 2005 quit state - sudden increase of brightness ($0.5^{m}$ in a few minutes) accompanied by increase of amplitude from $0.05^{m}$ to $0.2^{m}$ that never was detected after the outburst (Zharikov et al., 2008).  Despite J0804 was promissing CV for search for the WD non-radial g-mode pulsations, it did not show them before the 2006 outburst  (Szkody et al., 2006). We had pointed attention to the short-term light variations with most significant  12.6 min. oscillation  in 2008, January 1, when its amplitude at some pulses was commensurable with orbital light modulation $\sim 0.05^{m}$ (Pavlenko and Malanushenko, 2009).  Here the result of searching for the possible pulsations on the available base of our observations both before and after the 2006 outburst  in order to detect their first appearance is presented.

\section{Observations}
Taking into account the small amplitude of short-term variations of the SDSS J0804 the analysis was restricted by using of the most precise CCD data collected from 14 nights in 2006 - 2008 ensured accuracy of a single brightness measure better than $0.01^{m} - 0.15^{m}$ for time resolution of a few minutes for the multicolor observations and 20 - 30 seconds. They were obtained with 2.6-m Shajn telescope of the Crimean astrophysical observatory (CrAO) using the CCD FLI 1001E in Johnson B, V, R bands or in white light reduced to R. All measurements were made in respect to the comparison stars pointed by Zharikov et al (2008). The photometric standards in cluster M67 (Mendoza, 1967) were observed as well and used for measure of the V and R values of  comparison stars. Other details of observations are given in (Pavlenko and Malanushenko, 2009). The time resolution of observations was different and varied from a few minutes for multicolor observations to 15 - 20 seconds for observations in the white light.

\section{Outburst and orbital light curves}
The light curve of SDSS J0804 for 2006 - 2008 is given in Fig.1, including  one night   prior to the outburst. The dense part of points correspond to rebrightenings. After the end of rebrightenings the rapid brightness decline continued and has been observed during $\sim$ 3 - 4 weeks. Then the star became to fade very slowly.
\begin{figure}
\begin{center}
\includegraphics[width=36pc]{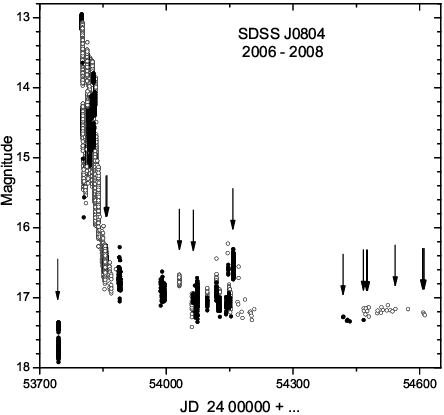}
\end{center}
\caption{\label{label}Light curve of the outburst of SDSS J0804. The V and R data are marked by filled and open circles respectively. Part of the data are averaged an expressed as one or two points per night. Arrows point to  the data used for the subsequent periodogram analysis.}
\end{figure}
We first found SDSS J0804 in the 2006 outburst (Pavlenko et al., 2007) at the end of the outburst plateau (two days before the plateau end). Assuming the possible plateau duration to be 2 - 3 weeks one could expect the start of the plateau outburst at JD $\sim$ 2453780 - 2453787.

The color-index V-R was estimated for several dates during the different stages of the outburst. For JD 2453856-2453858 (fast brightness fading after rebrightenings) the mean V = $16.4^{m}$,  V-R= $0.17^{m}$; for JD 2454479 (late fading stage) the mean V = $17.26^{m}$ and V-R = $0.11^{m}$.

The  dramatical change of the nightly light curve profile have been detected during the fast stage of the outburst fading in the three weeks after   rebrightenings were finished (see Fig. 2). To make sure that this modulation is really the orbital one, the data were folded on the orbital period using the ephemeris MinI = HJD 2453744.37 + 0.9590048(3)E  (Pavlenko and Malanushenko, 2009). One could see that  the light curve is already two-humped one, but the heights of the neighbor humps are $0.3^{m}$ and $0.05^{m}$ for (a), $0.3^{m}$ and $0.1^{m}$ for (b) and $0.2^{m}$ and $0.2^{m}$ for (c). Both minima are observed close to phases 0.0 and 0.5. While the slightly more deep minimum in (a) and (b) cases falls into phase 0.0, the more deep minimum in case (c) falls into the phase 0.5. Note the sharp eclipse-like minimum at phase near 0.0 for (a) and (b) cases.
The two-humped light curve and perfect fitting to the ephemeris are convincing arguments to regard this modulation as the orbital one.

The orbital light curves for more late stages of the outburst display small amplitude and show diversity of their on the whole two-humped profiles. In Fig. 3 the typical examples of the orbital light curves obtained in 195 - 775 days after the end of rebrightenings are shown. Some of them display two near-equal sine-like waves (a) per period, another -  WUMa-like neighbor humps (c) with round maxima and sharp minima. The curves (b) and (d) demonstrate rather four-humped light curve than two-humped one (or splitting of the humps with unequal heights).

The dependence of the amplitude of the orbital light curves on the mean brightness is shown in Fig. 4. The amplitudes  decrease  when brightness decrease  but with different rate. It is possible to pick out three different dependences: the first one corresponds to the data shown in Fig. 2 (they are  shown by the dashed line) - at that time the dependence is the sharpest, the second one belongs to the  data of the quiet long-lasting fading (dash-dotted line). Note that  the amplitude decreasing (expressed in intensities per unit of intensity)  was $\sim$ 10 times faster  during the fast outburst decline than during the slow fading. The third dependence corresponds to the data falling into the vicinity of maxima of the two mini-outbursts (dotted line). These amplitudes themself are slightly bigger than amplitude apart the mini-outbursts.
%\begin{figure}[h]
%\includegraphics[width=24pc]{lc_spjan08.eps}\hspace{2pc}%
%\begin{minipage}[b]{24pc}\caption{\label{label}}
%\end{minipage}
%\end{figure}

\begin{figure}[h]
\begin{minipage}{16pc}
\begin{center}
\includegraphics[width=12pc]{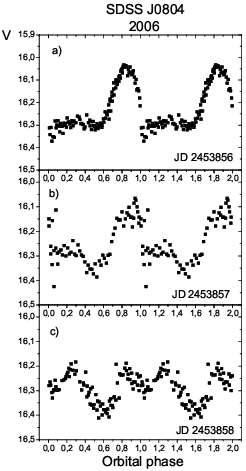}
\end{center}
\caption{\label{label}Example of data for April 30, May 01 and May 02, 2006  folded on the orbital period. For clarity data are plotted twice.}
\end{minipage}\hspace{2pc}%
\begin{minipage}{19.5pc}
\begin{center}
\includegraphics[width=21pc]{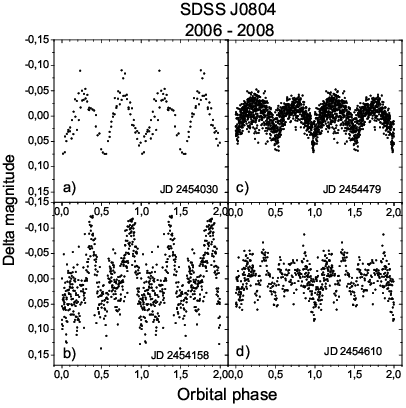}
\end{center}
\caption{\label{label}Example of data for October 21, 2006; February 26, 2007; January 13, 2008 and May 23, 2008 folded on the orbital period. For clarity data are plotted twice.}
\end{minipage}
\end{figure}

\begin{figure}
\begin{center}
\includegraphics[width=24pc]{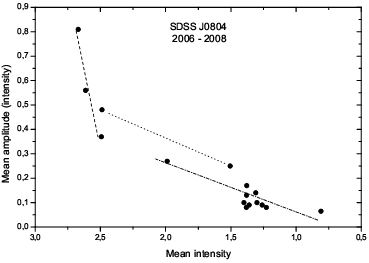}
\end{center}
\caption{\label{label}Dependence of the orbital amplitude on mean brightness expressed in the intensities (arbitrary  values). Dashed line is drawn through the data of April 30, May 01 and May 02 corresponding to the rapid outburst  decline after the end of  rebrightenings. Dash-dotted line is drawn through the data belong to the "`long tail"' of outburst and before one (last point), and dotted line is drawn through the data from two mini outbursts. }
\end{figure}

\section{Analysis of the short-term variations}
The white dwarf pulsations  in CVs is difficult to search for because these variations could be contaminated by other multiperiodical and quasi-periodical signals connected with orbital and rotational motion, instability in the accretion  stream and disk. The orbital light modulation profile itself also could vary from night to night as it was described above.
\begin{figure}
\begin{center}
\includegraphics[width=28pc]{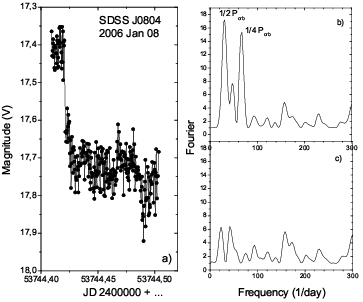}
\end{center}
\caption{\label{label}The light curve for the 2006, Jan 08 (a), Fourier transform for the initial data (b) and those for the data after orbital period removed (c).}
\end{figure}
The  investigation of the  short-term  light variations is performed for three data sets: 1) before outburst, 2)  during the fast outburst decline in $\sim$ 3 weeks after the end of rebrightenings and 3) during the long tail of the outburst.

The light curve for 2006, January 8 (1 - 1.5 months before outburst) is shown in Fig. 5 (a). One could see the jump-like brightness decrease from V = $17.4^{m}$ to V = $17.7^{m}$ occured during $\sim$ 9 minutes. The similar behavior had observed Szkody et al. (2006) also before outburst at JD 2453380  but in the "`opposite direction"': they detected the sharp brightness increase on $0.5^{m}$.  For the initial data from lower linear part of the light curve  the Fourier transform (FT) was calculated (see Fig.5 (b)). The most significant peaks point to the one quoter and one half of the orbital period. After the orbital wave subtraction the FT was calculated again for residuals. The ISDA package was used for calculations (Pelt, 1992). The amplitude periodogram is shown in Fig. 5 (c), where the amplitudes are normalized accordingly to Pelt (1992). No significant peaks were seen (Fig. 5 (c)) on that periodogram.

The mostly dense data from the second data set (JD 2453856) were selected for the analysis. The FT for original data is presented in Fig. 6 (a), where the significant peaks point to the orbital period and its twice value. FT for the data after orbital period removed is shown in Fig. 6 (b). There are a few peaks of the low significance, the most prominent one points to the frequency 86 $day^{-1}$ (16.7 min.). This corresponds to the light variations with mean amplitude of $0.025^{m}$ (see Fig. 6 (c)).
\begin{figure}
\begin{center}
\includegraphics[width=40pc]{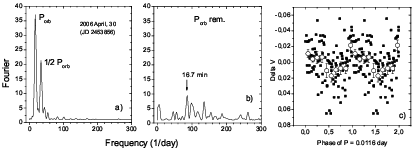}
\end{center}
\caption{\label{label}The Fourier transform for the initial data of 2006, April 30 (a), those for the data after orbital period removed (b) and data folded on the best period for residual spectrum 0.01156 day (16.7 min). For clarity data are plotted twice.}
\end{figure}
In Fig. 7 (a - k) the original light curves for the data from third set of observations are shown. For each data the FT after orbital period subtraction was calculated and correspondent periodograms are  performed in Fig. 7 (l - u). One could see that every periodogram (with exception  of (m)) shows a sequence of significant peaks. These series concentrate within 40 - 150 $day^{-1}$. The most stable pulsation corresponds to 114 $day^{-1}$ (12.6 min.) and its twice value 57 $day^{-1}$. The peak at 68 $day^{-1}$ could be caused by the four-humped structure of the orbital light curve profile because it coincides just with 1/4 $P_{orb}$. The less stable peak could be seen at frequency near 74 $day^{-1}$ and its twice value 148 $day^{-1}$. Many of peaks probably are not connected with WD pulsations.

\begin{figure}
\begin{center}
\includegraphics[width=36pc]{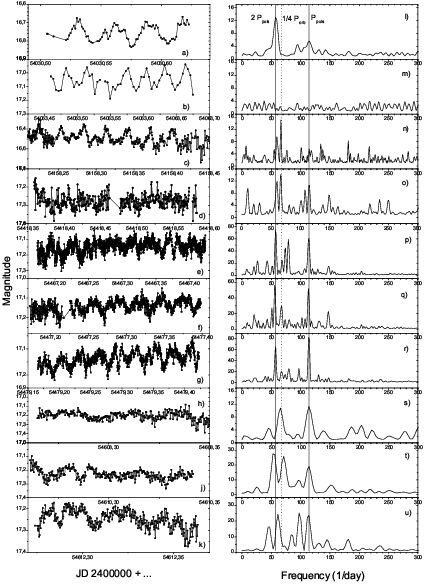}
\end{center}
\caption{\label{label}The light curves (a - k) and corresponding Fourier transforms (l - u) for the separated observations of SDSS J0804 in 2006 - 2008. The solid line are drown through the frequencies 57 and 114 $day^{-1}$ while the dotted one - through the 68 $day^{-1}$. }
\end{figure}

\section{Discussion}
The orbital light curves of SDSS J0804 having typically two-humped profile, often displayed the splitting of one or both humps.
The color-index V-R continued to decrease with time during the long brightness decay indicating to the decrease of the accretion disk contribution to the total light and, so, increase of the white dwarf contributing. The J0804 is localized in Galaxy at the low interstellar extinction, its E(V-R) could be calculated following Schlegel et al. (1998) and roughly is E(V-R) = $\sim 0.037^{m}$, so corrected value $(V-R)_{0}$ = $0.073^{m}$. That corresponds to the black-body temperature $\sim$ $15000^{\circ}$ K with accuracy of a few of thousands K. Taking into account that in $\sim$ 2 years after the outburst disk also could still contribute to the total light, one could expect that the temperature of the white dwarf  was hotter than $15000^{\circ}$ K.

The dependence of the orbital amplitude on the mean brightness could be caused by  the decreasing brightness of the bright detail on the accretion disk due to the decrease of the mass transfer rate over the source of outburst that occured $\sim$ 10 times faster during the fast outburst decline in respect to the late quiet state of slow outburst fading. Such dependence obviously is caused by decreasing initially more enhanced mass transfer rate, where decreasing itself fas $\sim$ 10 times faster  during the  fast outburst decline (after the end of rebrightenings) than during the late stage of outburst.

The variable and sometimes for-humped orbital modulation points to the complex and variable morfology of accretion disk.

We did not find any significant pulsations in 1 - 1.5 months before the outburst. The first detection of the most stable 12.6-min. pulsations was in $\sim$ 8 months after the expected start of the outburst. It is not clear whether the lack of pulsations at JD 2454063 was their temporary disappearance or caused by the insufficient data statistics. It is already known that the pulsators in dwarf novae could stop their pulsations by some reason  contrary to the ZZ Ceti stars (Southworth et al., 2008). So the lack of pulsations in SDSS J0804 in two occasions before outburst at JD 2453384 (Szkody et al., 2006) and  at JD 2453856 could not be the argument that before outburst the white dwarf never pulsated. However it is possible suggest that the compressional heating during the outburst and further fast cooling entered the white dwarf in the SDSS J0804 into the instability strip.

It is impossible also to consider confidently the 16.7 min. periodicity at JD 2453856 as the white dwarf pulsations because of the  periodogram is different from these performed in Fig. 7.
\subsection{Acknowledgments}
This work was partially supported by the grant F 25.2/139 of FRSF and CosmoMicrophysics program 5-20 of the Ukrainian National academy of sciense.
I am also grateful to the SOC and LOC  for the hospitality during the Conference.

\section*{References}
\begin{thereferences}
\item Mendoza E E. 1967 {\it BOOT} {\bf 4} 149
\item Pavlenko E P  et al. 2007 {\it 15th European Workshop On White Dwarfs} {\bf 372} Eds Ralf Napiwotzki and Matthew R. Burleigh. (San Francisko: ASP Conference Series) 511
\item Pavlenko E P. 2007 {\it Odessa Astronomical Publications} {\bf 20} 168
\item Pavlenko E P  and Malanushenko V P. 2009 {\it Kinematics and Physics of Celestial Bodies} {\bf 25} 72
\item Pelt Ja. 1992 {\it Irregulary Spaced Data Analysis, User Manual and Program Package}  (Helsinki: Helsinki University)  267
\item Schlegel D I Finkbeiner D P Davis M. 1998  {\it ApJ} {\bf 500} 525
\item Southworth J Townsley D M Gansicke B T. 2008 {\it MNRAS} {\bf 388} 709
\item Szkody P Henden A Agueros M et al. 2006 {\it AJ} {\bf 131} 973
\item Warner B Robinson E L. 1972 {\it Nature} {\bf 239} 2
\item Zharikov S V et al. 2008 {\it A\&A} {\bf 486} 505
\end{thereferences}
\end{document}